\documentstyle[preprint,aps]{revtex}
\newcommand{\be}{\begin{equation}}
\newcommand{\ee}{\end{equation}}
\newcommand{\bea}{\begin{eqnarray}}
\newcommand{\eea}{\end{eqnarray}}
\begin{document}
\draft
\title{\bf Classical and quantum mechanics of a particle on a
rotating loop}
\author{Sayan Kar \thanks{Electronic Address :
sayan@phy.iitkgp.ernet.in}} 
\address{Department of Physics and Centre for Theoretical Studies\\
Indian Institute of Technology, Kharagpur 721 302, INDIA}
\author{Avinash Khare \thanks{Electronic Address :
khare@iopb.res.in}} 
\address{Institute of Physics,\\
Sachivalaya Marg, Bhubaneswar 751 005, INDIA}
\maketitle
\parshape=1 0.75in 5.5in
\begin{abstract}
The toy model of a particle on a vertical rotating circle in the
presence of uniform gravitational/ magnetic fields is explored
in detail. After an analysis of the classical mechanics of the
problem we then discuss the quantum mechanics from both
exact and semi--classical standpoints. Exact solutions of the
Schr\"odinger equation are obtained in some cases by diverse
methods. Instantons, bounces are constructed and semi-classical,
leading order tunneling amplitudes/decay rates are written down. We also
investigate qualitatively the nature of small oscillations
about the kink/bounce solutions. Finally, the connections of
these toy examples with field theoretic and statistical
mechanical models of relevance
are pointed out.

\end{abstract}


\newpage

\section{Introduction}
Toy models in physics play an important role in
understanding the basic features of more involved theories
and phenomena. In particular, models in one dimensional
quantum mechanics, illustrating analogous situations in
a field theoretic context have been quite useful for
advanced researchers as well as beginning graduate
students. Among many such models, those which
illustrate non--perturbative aspects of field theory
through an analysis of instanton solutions and
bounces have been looked at in diverse contexts.
The $x^{2}-x^{4}$ potential and the quantum pendulum
have been discussed and analyzed in great detail
in the past. Apart from aspects which emerge out
of solutions and their analysis, it is also important
to relate the toy model with different
realistic models by mapping rules/dimensional extensions
etc. We shall, in this article, try to analyze a
novel toy model which has applications
in different situations and is also exactly
solvable to some extent. In particular, we discuss the
problem of  {\em Particle
on a Rotating Circle} (henceforth referred to as PORC) in the presence of 
gravitational/
magnetic field (a charged particle in the case of magnetic
field).

Before we indulge into analyzing the salient features of this model,
let us first explore the existing literature on it. For a
undergraduate/graduate student, a first
encounter with this model is likely to occur 
while doing a course in classical mechanics.
The model appears as a problem in the second chapter of
Goldstein's book on classical mechanics{\cite{gold:cm}}. 
It is also elaborately
discussed in Arnold {\cite{gold:cm}}. There exists a host of
other articles on it, in this journal too {\cite{porc:ajp}},
the most recent one being published a couple of years ago {\cite{flet:ajp97}}. 
The major directions along which this model has been
viewed are (i) as an example of spontaneous symmetry
breaking (ii) as a mechanical model for second order
phase transitions (iii) as an example in quantum mechanics 
where instanton solutions are obtainable {\cite{sk:pla92}}(iv) as a
lower dimensional analog of certain higher dimensional
field theories.

In this article, we shall see that this model (PORC) has the following
intriguing features :

\begin{itemize}
\item{The classical mechanics of the model is exactly
solvable in a way similar to that of the pendulum problem }
\item{When extended to full
real line (instead of circle $S^{1}$)\
 the model  has intimate connections with several
field theoretic and statistical mechanical models (for details see Section III).}
\item{The model in the presence of a uniform gravitational
or magnetic field has non-perturbative
solutions such as instantons/bounces and the corresponding
tunneling/decay rates can be written down and
analyzed (for a discussion on the theory of instantons, bounces and their
contribution to tunneling/decay rates can be found in \cite{gen:inst}, \cite
{gen1:inst}, \cite{kleinert:pi} ).} 
\item {A formal analysis of the Schr\"odinger equation can be performed
and exact solutions obtained. However the full expressions for the eigenvalues
and eigenfunctions are encoded in certain continued fractions which we do not
discuss in further detail.}
\item{In the presence of a uniform magnetic field the quantum problem is
 quasi-exactly solvable--we can get at least a few energy
eigenstates and eigenvalues 
analytically \cite{hks} \cite{km} by extending the method due to Razavy 
{\cite{raz:ajp}}.}
\end{itemize}

\section{The model}

Let us first introduce the model in some detail. We have a
point particle of mass $m$ (and charge $q$ if  we
are looking  at the problem in the presence of a magnetic field)
constrained to move on a
circle ($S^{1}$). The circle is located vertically and
is rotating at an angular frequency $\omega$ about the
vertical axis of symmetry. Two special cases will
be analyzed -- (i) the system in the presence of a
uniform gravitational field (ii) the system in the
presence of a uniform magnetic field. Specifications
about the directions and magnitudes of each of these uniform 
fields are given below. Note that this problem is a 
generalisation ($\omega \neq 0$) of the usual pendulum
problem ($\omega =0$) whose quantum mechanics is discussed
in {\cite{kp:ajp}}

A generic Lagrangian for the such a system can be given as :

\begin{equation}\label{2.1}
L_{generic} = \frac{1}{2}mr^{2}{\dot \theta}^{2} - A\cos \theta
- B\cos 2\theta \, ,
\end{equation}

where A and B are constants depending on the various parameters in
the model.

We can in principle have four cases depending upon the signs of A and B.
These are :

\centerline{(i) $A>0, B>0 $ (ii) $A>0, B<0$ (iii) $A<0,B>0$ (iv) $A<0, B<0$}

Note that with a proper redefinition of $\theta$ (more precisely
$\theta \rightarrow \pi - \theta$) we can relate the 
models (i) and (iii) as well as (ii) and (iv). Therefore, in essence we
have only two models to discuss. The two special cases given below
exemplify these two cases. 

{\bf Case 1 : System in a uniform gravitational field}

\begin{equation}
L_{grav} = \frac{1}{2}mr^{2}\dot \theta^{2} - \bigg (mgr \cos \theta - 
\frac{1}{2} mr^{2}
\omega^{2}\sin^{2}\theta \bigg ) \, , 
\end{equation}

where $\theta$ is the generalized coordinate required to
describe the system.

Thus the effective one--dimensional potential is :

\begin{equation}
V(\theta) = 
 mgr\cos\theta +\frac{1}{4}mr^{2}\omega^{2}\cos 2\theta - \frac{1}{4}
mr^{2}\omega^{2} \, . 
\end{equation}

Note that this system falls in the $A>0,B>0$ class mentioned above.
The location of the extrema of the effective potential are as follows :

{\bf Minima :}

(i) $\omega > \omega_{0}$ : $\theta = \cos^{-1}{(-a)} \, , 
2\pi - \cos^{-1}(-a) \, .$

$V(\theta_{min}) = V(2\pi - \theta_{min}) = -mr^2 \omega_0^2 (a^2+1)/a \, .$ 
  
(ii) $\omega <\omega_{0}$ : $\theta =\pi$ \hspace{.1in} ; \hspace{.1in}
 $V(\theta_{min}) = 
-mr^2 \omega_0^2 \, .$

{\bf Maxima :}

(i) $\omega >\omega_{0}$ : $\theta = 0(2\pi)$, $\theta =\pi$ \hspace{.1in} ;
\hspace{.1in}  
$V(0) = V(2\pi) = mr^2 \omega_0^2 \, ; \ V(\pi) = -mr^2\omega_0^2 \, .$

(ii) $\omega <\omega_{0}$ : $\theta = 0 (2\pi) \, .$
 
Thus for $\omega > \omega_0$ one has degenerate minima and maxima as well as a
local maxima while for $\omega < \omega_0$ there is no local maxima.
The potentials are shown in Fig. 1(a) and 1(b) for the $\omega>\omega_{0}$ and $\omega<\omega_{0}$
cases respectively. Here $\omega_0 = \sqrt{g/r}$ while $a = \omega_0^2/\omega^2$.

{\bf Case 2 : System in a uniform magnetic field} 

\begin{equation}
L_{mag} = \frac{1}{2}mr^{2}{\dot\theta}^{2} + \frac{1}{2}mr^{2}\omega^{2}
\sin^{2}\theta + q {\bf A . v} \, ,
\end{equation}
where $q$ denotes the charge of the particle.  

Since $\bf B$ is constant we have ${\bf A}= -\frac{1}{2} {\bf r \times B}$.
Assuming that the components of $\bf B$ are $(B_{r}, B_{\theta},0)$ we find
that the vector potential $\bf A$ has the components $(0,0,A_{\phi})$ where
$A_{\phi} = -\frac{1}{2} r B_{\theta}$. Since ${\bf v}$ has components 
$(0,r\dot\theta,
\omega r \sin \theta)$ we find that the Lagrangian takes the form :

\begin{equation}
L_{mag} = \frac{1}{2}mr^{2}{\dot \theta}^{2} - \left ( \frac{qBr^{2}\omega}{2} \sin
\theta - \frac{1}{2} mr^{2}\omega^{2} \sin^{2} \theta \right ) \, .
\end{equation}
Thus the effective one dimensional potential is
\begin{equation}
V (\theta ) = \frac{1}{2}mr^{2}\omega^{2}
 \left ( \frac{\omega_{c}}{\omega} \sin \theta -\sin^{2}\theta \right ) \, ,
\end{equation}
where $\omega_{c} = \frac{qB}{m}$.
With a straightforward redefinition of $\theta$ ($\theta \rightarrow
\theta - \frac{\pi}{2}$) we see that this model belongs 
to the other class ($A>0,B<0$).
The extrema of the effective potential 
are as follows :

{\bf Minima :} 

(i) $\omega > \omega_c/2 : \theta = \frac{\pi}{2} \, , \frac{3\pi}{2} \, ,$ 

$V (\frac{\pi}{2})
= \frac{1}{2}mr^{2}\omega^{2} \left ( \frac{\omega_{c}}{\omega} - 1 \right )
\quad ; \quad V (\frac{3\pi}{2}) = -\frac{1}{2}mr^{2}\omega^{2}
\left ( \frac{\omega_{c}}{\omega} + 1 \right )$

(ii) $\omega < \omega_c/2 : \theta = {3\pi\over 2} \quad ; 
\quad V(3\pi/2) = - {1\over 2} mr^2 \omega^2 ({\omega_c \over \omega} +1)$ 

{\bf Maxima :}

(i) $\omega > \omega_c/2 : \theta = \sin^{-1}\frac{\omega_{c}}{2\omega},
\ \pi - \sin^{-1}\frac{\omega_{c}
}{2\omega} \quad ; \quad V (\theta_{max}) = \frac{1}{8}mr^{2} \omega_{c}
^{2}$ \, .

(ii) $\omega < \omega_c/2 : \theta = {\pi \over 2} \quad ; \quad 
V(\pi/2) = {1\over 2} mr^2 \omega^2 ({\omega_c \over \omega} -1)$ \, .

Thus for $\omega > \omega_c /2$ the effective potential represents a 
system which has a false
vacuum at $\theta = \frac{\pi}{2}$ and a true vacuum at 
$\theta = \frac{3\pi}{2}$ while for $\omega < \omega_c /2$ there is no local 
minima.
The two scenarios
are plotted in Fig. 2(a)  and  2(b) respectively.

>From the figures for the effective potentials we can conclude
the following -- (i) in the presence of a uniform magnetic field we may have 
a true as well as false vacua  
while (ii) in  a uniform gravitational field we can have a degenerate
double well potential. Additionally, if we generalize the 
above over the full (real)
line then we have  
a periodic potential for which  
the minima are infinite-fold degenerate.

A more general problem is that of a charged particle in a uniform gravitational 
plus magnetic (and even electric) field. The most general Lagrangian is given
by 
\begin{equation}\label{2.7}
L_{general} = \frac{1}{2}mr^{2}{\dot \theta}^{2} +\frac{1}{2}mr^{2}
\omega^{2} \sin^{2} \theta - mgr\cos\theta -\frac{qBr^{2}\omega}{2}
\sin \theta + q\phi_{elec} \, ,
\end{equation}
where $\phi_{elec}$ is the electric potential. In general, this problem
cannot be exactly solved in either classical or quantum mechanics even
though a general analysis of the motion in classical mechanics is 
possible {\cite{flet:ajp97}}. 
However, by suitably choosing 
the ratio of the electric and magnetic
fields we can reduce the problem to that of a particle
moving on a rotating circle in uniform gravitational
field alone. 
 Similarly, by suitably choosing the ratio of the gravitational and
electric fields, one can reduce it to that of a particle moving on a 
rotating circle in a uniform magnetic field alone. 
Other combinations and their consequences can also be tried out. We leave it
to the reader to figure out the details of these scenarios.

\section{Inter--connections with field theoretic and other models}

The toy models discussed above have remarkable connections
with a wide--ranging variety of field theoretic models. We shall now 
briefly summarize some of these interconnections.

To begin with, note that there exists a correspondence between the toy
models discussed above and $1+1$ dimensional double-- sine--Gordon (DSG) 
field theory defined by the Lagrangian density : 
\begin{equation}\label{3.1}
{\cal L} = {1\over 2}{\partial_{\mu} \phi} {\partial^{\nu}\phi} + Acos \phi 
+ B cos 2\phi \, ,
\end{equation}
in that the potential is 
formally the same in both the cases.  
Here A and B could be positive or negative. 
The field equation that follows 
from a first variation is  
\begin{equation}\label{3.2}
{\partial^2 \phi \over \partial t^2} - {\partial^2 \phi \over \partial x^2} = A \sin \phi + 2B \sin 2\phi \, .
\end{equation}
Hence the equation for static (i.e.  
$\phi$ independent of time) solutions is  
\begin{equation}\label{3.3}
{d^2 \phi \over dx^2} = -\left ( A sin \theta + 2B sin 2\theta \right ) \, .
\end{equation}
It may be noted that this equation also 
follows  from our toy model Lagrangian as given by eq. (\ref{2.1}) but for
an overall sign. In fact if we go to the Euclidean time i.e. let $t = i\tau$
, then the classical equation of motion in the toy model takes the form
\begin{equation}\label{3.4}
{d^2 \theta \over d\tau^2} = - (A  sin \theta + 2B  sin \theta) \, ,
\end{equation}
which is identical to the static field equation as given by eq. (\ref{3.3}).
Now it is well known {\cite{gen1:inst}} that the finite Euclidean 
action solutions of the 
Euclidean equations of motion (\ref{3.4}) for any system  are nothing but the 
instantons,  whose mere existence is due to the presence of degenerate
minima in the potential appearing in the theory. Further, once one has 
obtained  
these instanton solutions and
the corresponding Euclidean actions, the tunneling 
amplitude for the transition 
between the degenerate vacua can easily be computed in the dilute
(non-interacting) instanton  gas
approximation. We thus notice that the static finite energy solutions of 
the DSG equation as given by (\ref{3.3}) are the same as the finite 
(Euclidean) action instanton solutions of our Toy model (with the obvious
replacement of $\phi$  with $\theta$ and $x$ with $\tau$). 

Since the DSG
equation has been extensively studied in the literature 
{\cite{leu,ptk,boc,cgm,det}},
we can immediately write down the corresponding 
instanton solutions of our toy model.

It may be added here that the DSG field equation 
occurs quite naturally in several 
physically important  situations. For example, it appears
quite naturally in the study of spin waves in the B-phase of super-fluid $^3 He$
\cite{mat,buc}, in the problem of self-induced transparency 
\cite{dbc} and in nonlinear excitations in a compressible chain of XY dipoles
under 
conditions of piezoelectric coupling \cite{rem}. Further, a  model based on 
the DSG theory  
has been proposed \cite{hls} to describe the poling process in the 
B-phase of the 
polymer polyvinylidene fluoride ($PVF_2$). Additionally, the DSG theory 
has been investigated as a 
model of a nonlinear system which can support more than one
kind of soliton. 

The sine-Gordon equation, in turn, is related to
the massive Thirring Model \cite{col} while a specific form of DSG has been
shown to be equivalent to a generalized massive Thirring model \cite{bha}. 
Thus, any new results obtained in any of these models will immediately have
relevance in various other related models.

Finally, we may add that the DSG equation is also related to the 
anisotropic Heisenberg chain in an applied
magnetic field. In particular, by treating spins classically, Leung \cite{leu} 
as well as  
Pandit et al. \cite{ptk} have shown that the  
classical spin dynamics is approximately governed by the generalized double
sine Gordon (DSG) model as given by eq. (\ref{2.7}). 

\section{Classical Solutions}

We now move on towards analyzing the real time solutions i.e. the
classical mechanics of the above problem. 
Defining $E=T+V-V_{min}$ so that $E \ge 0$, 
we integrate the corresponding quadrature
and then analyze the various cases separately.

{\bf 1. System in a gravitational field}

For the gravitational case, for the two domains of
$\omega$ we have the following results.. 

{\underline{\sf $\omega <\omega_{0}$}}

Here $V_{min}=-mr^{2}\omega_{0}^{2}$. Using $\theta = 2\phi$ and
then $\tan \phi = y$ we obtain the following generic form of integral which we
need to analyze :
\begin{equation}
\int \frac{dy}{\sqrt{y^{4} + 2y^{2}\left (1-\alpha -\beta \right )
+ \left (1-2\alpha\right )}} = \sqrt{\frac{E}{2mr^{2}}} \left ( t-t_{0}\right )
\end{equation}
where
\begin{equation}
\alpha = {m\omega_0^2 r^2\over E} \quad ; \quad 
\beta = {m\omega^2 r^2 \over E} \, .
\end{equation}
Recall that $V_{max}=2mgr=2mr^{2}\omega_{0}^{2}$. We can therefore
have two possibilities $E<V_{max}$ and $E>V_{max}$ which we discuss one by one.

{\sf (a)  $E<V_{max}$}

In this energy range we have the following
two alternatives depending on the sign of the quadratic term
in the denominator of the elliptic integral. These are :

(i) $0<E<mr^{2}\left (\omega_{0}^{2}-\omega^{2} \right )$

In this case the solution is \cite{gr}  
\begin{equation}
\tan \frac{\theta}{2} = b \frac{dn(X,k)}{\kappa^{\prime}sn(X,k)}
\end{equation}
where $X=\frac{b}{\kappa^{\prime}}\sqrt{\frac{E}{2mr^{2}}}(t-t_{0})$,
$a^{2}b^{2}=2\alpha -1$, $ b^{2}-a^{2} = 2 (\alpha -\beta -1)$ \, .
Here cn(X,k), sn(X,k) and dn(X,k) are Jacobi elliptic functions of real 
elliptic modulus parameter $k$. 

(ii) $mr^{2}\left (\omega_{0}^{2}-\omega^{2} \right ) < E <V_{max}$

In this case the solution is 
\begin{equation}
\tan {\frac{\theta}{2}} = \frac{a dn(X,k)}{\kappa sn(X,k)}
\end{equation}
where $X=\frac{a}{\kappa}\sqrt{\frac{E}{2mr^{2}}}(t-t_{0})$, $ \frac{a}{a^{2}+b^{2}}
=\kappa$, $\kappa^{\prime} = \sqrt{1-\kappa^{2}}$, 
$a^{2}b^{2} = 2\alpha -1$, $a^{2}-b^{2}=2(1-\alpha+\beta)$.

{\sf (b) $E>V_{max}$}

The exact solution here is :
\begin{equation}
\tan \frac{\theta}{2} = a cn(X,k)/sn(X,k)
\end{equation}
where $X= a\sqrt{\frac{E}{2mr^{2}}}(t-t_{0})$, $\frac{\sqrt{a^{2}-b^{2}}}{a}
=\kappa$, 
$a^{2}b^{2}=1-2\alpha$, $a^{2}+b^{2}=2(1-\alpha+\beta)$.
Note also that as $b\rightarrow 0$, $a^{2}=1+\frac{\omega^{2}}{\omega_{0}^{2}}$
, which implies $E=2mr^{2}\omega_{0}^{2}$.

{\underline{\sf $\omega >\omega_{0}$}}

In this domain of $\omega$ we note that $V_{min}= 
-\frac{mr^{2}}{2\omega_{0}^{2}} \left (\omega_{0}^{4} +\omega^{4} \right )
$. As before, using $\theta = 2\phi$, $\tan \phi = y$ we get :
\begin{eqnarray}
\int \frac{dy}{\sqrt{y^{4} \left ( E-V_{max} + 2mr^{2}\omega_{0}^{2} \right )
+ 2y^{2} \left ( E + mr^{2}\omega_{0}^{2} + mr^{2}\omega^{2} 
-V_{max} \right ) + \left ( E-V_{max} \right ) }} \nonumber \\
= \sqrt{\frac{1}{2mr^{2}}}
\left (t-t_{0} \right ) \, ,
\end{eqnarray}
where $V_{max} = \frac{m r^2}{2 \omega^2} \left (\omega_{0}^{2} + \omega^{2}
\right )^{2}$. The exact solutions in the various sub-cases are analyzed below.

{\sf (a) $E<V_{max}$}

In a way similar to the $\omega<\omega_{0}$ case we once again have two
possibilities depending on the signs of the coefficients of the various
terms appearing in the elliptic integral.

(i)  $0<E<\frac{mr^{2}}{2\omega^{2}}\left (\omega^{2}-\omega_{0}^{2}
\right )^{2}$

The solution for this case is :
\begin{equation}
\tan \frac{\theta}{2} = \frac{1}{a dn(X,k)}
\end{equation}

where $X= a\sqrt{\frac{V_{max}-E}{2mr^{2}}} (t-t_{0})$, $a^{2}b^{2} =
\frac{V_{max} - E - 2mr^{2}\omega_{0}^{2}}{V_{max} - E}$,
$a^{2} + b^{2} = 2\frac{E-V_{max}+mr^{2}(\omega^{2}+\omega_{0}^{2})}
{V_{max} - E}$.

(ii)  $\frac{mr^{2}}{2\omega^{2}}\left (\omega^{2}-\omega_{0}^{2}
\right )^{2}< E <V_{max}$

Within these bounds of energy, the solution turns out to be 

\begin{equation}
\tan {\frac{\theta}{2}}=  \frac{a}{\kappa} \frac{dn(X,k)}{sn(X,k)}
\end{equation}

with $X= \frac{a}{\kappa} \sqrt{\frac{E+2mr^{2}\omega_{0}^{2} - V_{max}}
{2mr^{2}}} (t-t_{0})$, $a^{2}b^{2} = \frac{V_{max}-E}{E+2mr^{2}\omega_{0}^{2}
-V_{max}}$, $a^{2}-b^{2} = \frac{E+mr^{2}(\omega^{2}+\omega_{0}^{2})-V_{max}}
{E+2mr^{2}\omega_{0}^{2} -V_{max}}$.

{\sf (b) $E>V_{max}$}

The solution here is given by:

\begin{equation}
\tan{\frac{\theta}{2}} = a\frac{cn(X,k)}{sn(X,k)}
\end{equation}
where $X=a\sqrt{\frac{E-V_{max} +2mr^{2}\omega_{0}^{2}}{2mr^{2}}} (t-t_{0})$.

{\bf 2. System in a magnetic field :}

In the magnetic case, we follow the same procedure as above. 
The only major difference is that, unlike the gravitational case, $V_{min} =
-\frac{1}{2} mr^{2}\omega^{2} \left ( 1 +\frac{\omega_{c}}{\omega} \right )$
is the same irrespective of whether 
 $\omega>\frac{\omega_{c}}{2}$ or $\omega < \frac{\omega_{c}}{2}$. 
Using $\theta = \phi -\frac{\pi}{2}$, $\frac{\phi}{2} = \eta$ and $\tan \eta
=y$ we find that we have to handle the following integral :
\begin{equation}
\int \frac{dy}{\sqrt{\left [ \left (1-\alpha\right ) y^{4} + \left (2-\alpha
-\beta \right ) y^{2} + 1 \right ]}} = (t-t_0) \sqrt{E\over 2mr^2} \, ,
\end{equation}
where $\alpha = {mr^2 \omega \omega_c \over E}, \beta = {2mr^2 \omega^2
\over E}$.
 
It is convenient to make another transformation $y=\frac{1}{u}$ and then
analyse the resulting integral, which is, generically, of the form 

\begin{equation}
-\int \frac{du}{\left [ u^{4} + u^{2} \left ( 2-\alpha -\beta \right)
 + \left ( 1-\alpha \right
) \right ]^{\frac{1}{2}}} = 
\left (t-t_{0}\right ) \sqrt{\frac{E}{2mr^{2}}} \, .
\end{equation}
The various cases are now analyzed below.

{\underline{\sf $\omega < \frac{\omega_{c}}{2}$}}

{\sf (a) $E<V_{max}$}

Here, depending on the signs of the various coefficients we will have
two cases :

(i) $\frac{mr^{2}\omega\omega_{c}}{2}\left (1+\frac{2\omega}{\omega_{c}}
\right ) < E < V_{max}$

The solution here is :
\begin{equation}
\tan \left (\frac{\theta}{2}
+\frac{\pi}{4} \right ) 
= \frac{k}{a} \frac{sn (X,k)}{dn(X,k)}
\end{equation}
where $\kappa = \frac{a}{\sqrt{a^{2}+b^{2}}}$, $a^{2}b^{2}=\alpha-1$,
$a^{2}-b^{2}=2-\alpha-\beta$ and $X= \frac{a}{\kappa}\sqrt{\frac{E}{2mr^{2}}}
(t-t_{0})$.

(ii) $0<E<\frac{mr^{2}\omega\omega_{c}}{2}\left (1+\frac{2\omega}{\omega_{c}}
\right )$

The solution here is the same as before except that now 
 $a^{2}<b^{2}$ and therefore
one may obtain it from the previous solution by just interchanging
$a$ and $b$. 

{\sf (b) $E > V_{max}$}

\begin{equation}
\tan \left ( \frac{\theta}{2} +\frac{\pi}{4} \right ) = \frac{sn(
X,k)}{acn(X,k)}
\end{equation}

 where $ X=a{\sqrt{E \over 2mr^{2}}} (t-t_0)$,
$a^2 b^2 = 1-\alpha$, $a^2+b^2 =2-\alpha-\beta$ \, .

{\underline{\sf $\omega > \frac{\omega_{c}}{2}$}}
 
{\sf (a) $E<V_{max}$}

The two cases depending on the signs of the various coefficients are :

(i) $0<E<mr^{2}\omega\omega_{c}$
Here the solution is :

\begin{equation}
\tan \left (\frac{\theta}{2}+\frac{\pi}{4} \right )
 = \frac{k'^2}{b} \frac{sn(X,k)}{dn(X,k)} \, , 
\end{equation}
where $a^{2}b^{2} = \alpha - 1$, $b^{2}-a^{2}=\alpha +\beta -2$, $\kappa
=\frac{a}{\sqrt{a^{2}+b^{2}}}$ and of course $k'^2+k^2 =1$.

(ii) $mr^{2}\omega\omega_{c}<E<V_{max}$

The solution is 
\begin{equation}
\tan \left (\frac{\theta}{2} + \frac{\pi}{4} \right ) = \frac{1}{a}
sn [a\sqrt{\frac{E}{2mr^{2}}} (t-t_{0} )] \, .
\end{equation}

{\sf (b) $E>V_{max}$}

The two cases are given below :

(i) $V_{max} < E < mr^{2}\left (\omega^{2}+\frac{\omega\omega_{c}}{2}\right )$

The solution is 

\begin{equation}
\tan \left (\frac{\theta}{2}+\frac{\pi}{4} \right ) = \frac{1}{a}
sn(X,k) \, .
\end{equation}

(ii) $E>mr^{2}\left (\omega^{2} + \frac{\omega\omega_{c}}{2} \right )$

The solution is 

\begin{equation}
\tan \left (\frac{\theta}{2} + \frac{\pi}{4} \right ) = \frac{1}{a}
\frac{sn(X,k)}{cn(X,k)} \, ,
\end{equation}
where in both the above solutions $X=a\sqrt{\frac{E}{2mr^{2}}} (t-t_{0})$.

Before concluding this section, we briefly discuss 
a couple of special cases in which the solution can be written in terms of
the well known trigonometric and hyperbolic functions.

{\underline{\sf Special solutions in the gravitational case}

For $\omega < \omega_{0}$ if we choose $E= 2mr^{2}\omega_0^{2}$ then we 
obtain the following  special solution  
\begin{equation}
\tan \frac{\theta}{2} = \pm\frac{1}{a} cosech \left (\sqrt{\omega_{0}^{2}
+ \omega^{2}} (t-t_0)\right ) \, ,
\end{equation}
where $a^2 = \frac{\omega_{0}^{2}}{\omega^{2}+\omega_{0}^{2}}$.
This is known as the `sticking' solution--i.e. as $t\rightarrow \pm \infty,$
$\theta \rightarrow 0$ (maxima), while for $t\rightarrow t_{0},$ 
$\theta \rightarrow \pi$ (minima).

If $\omega> \omega_{0}$, the same solution holds (with the same form of $a$)
but at $E = V_{max} = {mr^2 \over 2\omega^2} (\omega^2 +\omega_0^2)^2$.
It is also a sticking solution but it goes from local to absolute maxima
as $t$ goes from $t_0$ to $\pm \infty$.
It is the {\em value} of $\omega$ which makes the two solutions functionally
different. 

{\underline{\sf Special solutions in the magnetic case}

Here, for $\omega<\frac{\omega_{c}}{2}$ and $E = V_{max} 
=mr^{2}\omega\omega_{c}$
we find the following solution :

\begin{equation}
\tan ({\frac{\theta}{2} +\frac{\pi}{4}}) = \sqrt{\frac{1}{1-\frac{2\omega}
{\omega_{c}}}} \sinh \left (\sqrt{\frac{\omega\omega_{c}}{2} 
 \left (1-\frac{2\omega}{\omega_{c}} \right ) }(t-t_0)\right )
\end{equation}
This is once again the so--called `sticking' solution--the particle
reaches the maxima ($\pi /2$) as $t\rightarrow \pm \infty$ while it is at the
minimum ($3\pi /2$) as $t\rightarrow t_0$.

Another solution for $\omega<\frac{\omega_{c}}{2}$ is obtained
at $ E=\frac{mr^{2}}{8} \left (\omega_{c} + 2\omega \right )^{2}
 > V_{max}$.
This is given by 
\begin{equation}
\tan \left ( \frac{\theta}{2} +\frac{\pi}{4} \right ) = \sqrt{\frac{
\omega_{c} +2\omega}{\omega_{c} - 2\omega}}
\tan \left (\sqrt{1-\frac{4\omega^{2}}{\omega_{c}^{2}}} \frac{\omega_{c}}{4}
(t-t_0) \right ) \, .
\end{equation}
This solution oscillates around the minima at $\theta =\frac
{3\pi}{2}$.

For $\omega>\frac{\omega_{c}}{2}$ and $E=mr^{2}\omega\omega_{c} < 
V_{max}$ we have
a solution which oscillates around the minimum at $\theta=\frac{3\pi}{2}$.
This is given by 
\begin{equation}
y=\tan \left (\frac{\theta}{2} + \frac{\pi}{4} \right ) =
\sqrt{\frac{\omega_{c}}{2\omega-\omega_{c}}} \sin \left (\sqrt{1-
\frac{\omega_{c}}{2\omega}} \omega (t-t_{0}) \right ) \, .
\end{equation}
Finally, for $\omega > {\omega_c\over 2}$ and $E = {mr^2 \over 8}
(\omega_c +2\omega)^2 = V_{max}$ we have the solution
\begin{equation}
\tan \left ( \frac{\theta}{2} +\frac{\pi}{4} \right ) = \sqrt{\frac{
2\omega +\omega_{c}}{2\omega - \omega_{c}}}
\tanh \left (\sqrt{1-\frac{\omega_{c}^{2}}{4\omega^{2}}} \omega
(t-t_0) \right ) \, .
\end{equation}
This is again a sticking solution which goes from the absolute minimum to
the maximum as $t$ goes from $t_0$ to $\pm \infty$.
 
\section{Quantum mechanics of a PORC 
in a constant gravitational field--instantons}

In the gravitational case, the potential has 
a pair of degenerate minima. Therefore, one has the
possibility of constructing instanton solutions. Since the
two minima are separated by two different barriers there are
two types of instantons. In Appendix we also look at the possibility of
constructing
exact solutions of the Schr\"odinger equation in this case.  

\subsection{Instanton solutions}

As mentioned earlier, the effective potential has degenerate
minima and therefore we will obviously have instanton solutions.
Since the minima are separated by two kinds of barriers we will
have two different instantons --one across the barrier at $0 (2\pi)$
and another across the barrier at $\pi$. 

It is worth reminding that instantons are the finite Euclidean 
action solutions of the Euclidean equations of motion of
any theory. Since we are dealing with particle mechanics we
look at solutions to the Euclideanised Newton's second law.
These solutions have the interpretation in terms of 
quantum tunneling. For a particle in a double well potential, classically,
it is not possible to cross the barrier connecting the two
vacua. However, when we Euclideanise the Newton's second law, we are
essentially looking at the motion in Euclidean time in an inverted potential.
These Euclidean solutions (called instantons) begin at one vacuum at
$\tau\rightarrow -\infty$ and end at another at $\tau\rightarrow +\infty$.
In the saddle--point approximation,
the tunneling amplitude goes as 
$\exp(\frac{-S_{E}}{\hbar})$.

The instanton solutions and aspects of quantum tunneling in this model
have been worked out in an earlier paper by one of us  
\cite{sk:pla92}. We
therefore only summarize these results here and refer the reader to that
article for the relevant details. This section is included
here entirely for the sake of completeness.

Below we write down the instanton solutions and the 
corresponding Euclidean actions. To arrive at the instantons
we need to solve the Euclidean equation of motion which is 
given by
\begin{equation}
\theta '' = - \omega^{2} (\cos \theta + a) \sin \theta \quad ; \quad 
a = {\omega_0^2 \over \omega^2} \, ,
\end{equation} 
where the prime denotes differentiation w.r.t. $\tau = -it$ (Euclidean time).
It is easy to convert the above equation into one for
$\theta '$ alone and then integrate the resulting first order
equation.

We first deal with the case of $\omega >\omega_{0}$.
The  instanton/anti--instanton and it's Euclidean action across the barrier 
at $\theta=2\pi(0)$ are given by 
\be\label{t1}
\theta_{1}(\tau) = \pm 2\tan^{-1}\left ( \sqrt{\frac{1+a}{1-a}}\tanh\left [
\sqrt{\frac{1}{4}(1-a^{2})}\omega\tau\right ]\right ) \, .
\ee
\be
S_{E1}=4mg^{1/2}r^{3/2}\left [ \sqrt{\frac{1-a^{2}}{a}}+2\sqrt{a} \tan^{-1}
\left (\sqrt{\frac{1+a}{1-a}} \right ) \right ] \, .
\ee

The instanton/anti--instanton
and it's Euclidean action across the barrier at $\theta=\pi$ are given by :
\be\label{t2}
\theta_{2}(\tau) = \pm 2\tan^{-1}\left ( \sqrt{\frac{1+a}{1-a}}\coth\left [
\sqrt{\frac{1}{4}(1-a^{2})}\omega\tau\right ]\right ) \, .
\ee
\begin{equation}
S_{E2}=4mg^{1/2}r^{3/2}\left [ \sqrt{\frac{1-a^{2}}{a}}-2\sqrt{a} \tan^{-1}
\left (\sqrt{\frac{1+a}{1-a}} \right ) \right ] \, .
\end{equation}

Notice that for $a\sim 1$ (i.e. $\omega \sim \omega^{0}$) 
the second term in each Euclidean action has the larger
value. This implies that the Euclidean action is proportional to
$\pm \frac{4\pi mgr}{\hbar \omega}$ and the tunneling
amplitude will be dominated by the instanton across the $\theta =\pi$
barrier. On the other hand for $a\sim 0$ (i.e. $\omega \sim \infty
$ or $r \rightarrow \infty $--very high frequencies or very large
radius ) the Euclidean action is proportional to $\frac{4m\omega^{2}r}{\bar h
\omega}$ and the contribution to the tunneling probablity from both
the instantons are the same. In the latter case, the effect of
gravity is washed out while in the former gravity predominates.
 
For the $\omega <\omega_{0}$ case we have a single well in $S^{1}$ but 
infinite number of degenerate minima on the full line and the instanton/
anti--instanton as well as it's Euclidean action are given by
\be\label{t3}
\theta_{3} (\tau) = \pm 2\tan^{-1}\left ( p \sinh (\omega_0 \tau \over p) 
\right ) \, .
\ee
\begin{equation}
S_{E3} =4mg^{1/2}r^{3/2}\left [ \sqrt{a}\tan^{-1}\left (\frac{1}{\sqrt{
a-1}} \right ) + \sqrt{\frac{a-1}{a}} \right ]
\end{equation}
with $p =\sqrt{\frac{a}{a-1}}$.

For domains of $a\sim 1$ (when the first term in $S_{E3}$ dominates
) and $a >>1$ (when the second term in $S_{E3}$ gives the largest
contribution)  one can see the effects similar to 
the ones stated earlier. 

\section{Quantum mechanics of a PORC in a constant magnetic field
--bounces, instantons and exact solutions}

\subsection{ Exact Solutions} 

We now write down the Schr\"odinger equation for this system and look
for exact solutions \cite{hks} following the method introduced by Razavy for
bistable potentials {\cite{raz:ajp}}.

For this case, the Schr\"odinger equation turns out to be 
\begin{equation}\label{se}
\frac{d^{2}\Psi}{dx^{2}} + \left [ \frac{8mr^{2}E}{{\hbar}^{2}} + 
\frac{2m^{2}r^{4}\omega^{2}}{{\hbar}^{2}} - \frac{4m^{2}r^{4}\omega
\omega_{c}}{{\hbar}^{2}} \cos 2x + \frac{2m^{2}r^{4} \omega^{2}}
{{\hbar}^{2}} \cos 4x \right ] \Psi = 0 \, ,
\end{equation}

where we have introduced the following redefinitions $\theta --> \frac{\pi}{2}
- \theta$ and then $\theta --> 2 x$.
We further rewrite the Schr\"odinger equation in the following form :
\begin{equation}
\frac{d^{2} \Psi}{dx^{2}} + \left [ \bar \epsilon 
-(n+1)\xi \cos 2x + \frac{1}{8} \xi^{2} \cos 4 x \right ] \Psi = 0
\end{equation}
where 
$ \epsilon = \frac{8mr^{2}E}{\hbar^{2}} \quad ; \quad \bar \epsilon
= \epsilon + \frac{1}{8}\xi^{2} \quad ; \quad 
(n+1) = \frac{mr^{2}\omega_{c}}{\hbar} \quad ; \quad 
\xi = {4m r^2 \omega \over \hbar}$ .

We shall now write down the exact solutions for different values of
$n$ which, of course, correspond to different Lagrangians/Hamiltonians
with the same functional form. In fact, as is clear from the
definition given above,  $n$ is related to the ratio of the {\em critical}
angular momentum $mr^{2}\omega_{c}$ and Planck's constant
$\hbar$. 

Following Razavy, we first write down a solution for $n=0$
(and also $\bar \epsilon =0$) and then
define the wave functions for $n>0$ as the product of the $n=0$
solution and an unknown function $\Phi (x)$.
We thus start with the ansatz
\begin{equation}
\Psi (x) = \exp \left ( -\frac{\xi}{4}\cos 2x \right ) \Phi (x) \, . 
\end{equation}
The resulting differential equation for $\Phi (x)$ turns out to be 
\begin{equation}
\Phi^{''} + \xi \sin 2x \Phi^{'} + \left ( \bar \epsilon + {\xi^2 \over 8} 
- n \xi \cos
2x \right ) \Phi = 0
\end{equation}
with the primes denoting differentiation w.r.t. $x$.
It is easily shown that this is a quasi-exactly solvable (QES) system 
\cite{km}. In
particular, one can easily show that the above equation is equivalent to 
a three term difference equation which is exactly solvable in case $n$ is a 
non-negative integer. In particular, for $n= 0,1,2,...$, first $(n+1)$ 
solutions of period $\pi$ ($2\pi$) are obtained in case $n$ is even (odd).
For example, the solutions for $n= 0,1,2,3$ are  

${\bf n = 0 :}$

\centerline{ $ \Phi_{0} = 1 \quad ; \quad $  $E_{0}
= -{1\over2}\hbar \omega \frac{\omega}{\omega_{c}}$} \, .
  
${\bf n = 1 :}$

\centerline{$\Phi = \sin x \quad ; \quad $  $
E = \frac{1}{16}\hbar \omega_{c} - \frac{1}{2} \hbar \omega
\left ( \frac{1 + 2\omega}{\omega_{c}} \right ) $
} \, .

\centerline{$\Phi = \cos x \quad ; \quad $  $
E = \frac{1}{16}\hbar \omega_{c}  + \frac{1}{2} \hbar \omega
\left (1-2\frac{\omega}{\omega_{c}} \right ) $} \, .

${\bf n = 2 :} $

\centerline{ $\Phi_{0} = 2\xi +a_{-} \cos 2x \quad ; \quad $
$ E_{0} = \frac{\hbar \omega_{c}}{12} -\frac{3}{2}
\hbar \omega \frac{\omega}{\omega_{c}} - \frac{\hbar\omega_{c}}{12}
\sqrt{1 + 144\frac{\omega^{2}}{\omega_{c}^{2}}}$
} \, .
\centerline{$\Phi = \sin 2x \quad ; \quad $
$E = \frac{1}{6}\hbar \omega_{c} - \frac{3}{2}
\hbar \omega \frac{\omega}{\omega_{c}}$
} \, .
 
\centerline{$\Phi = 2\xi +a_{+}\cos 2x \quad ; \quad $
$ E = \frac{\hbar \omega_{c}}{12} -\frac{3}{2}
\hbar \omega \frac{\omega}{\omega_{c}} + \frac{\hbar\omega_{c}}{12}
\sqrt{1 + 144\frac{\omega^{2}}{\omega_{c}^{2}}}$} \, .

$\bf n = 3: $

\centerline{$\Phi = 3\xi \cos x + b_{0}\cos 3x \quad ; \quad $
$E = \frac{\hbar \omega_{c}}{32} \left ( -\frac{1}{4}\xi^{2} + 5
+ \xi -2 \sqrt{\xi^{2} +4 - 2\xi} \right )$} \, .

\centerline{$\Phi = 3\xi \sin x + b_{1}\sin 3x \quad ; \quad $
$E = \frac{\hbar \omega_{c}}{32} \left ( -\frac{1}{4}\xi^{2} + 5
- \xi -2 \sqrt{\xi^{2} +4 + 2\xi} \right )$} \, .

\centerline{$\Phi = 3\xi \cos x + b_{2}\cos 3x \quad ; \quad $
$E = \frac{\hbar \omega_{c}}{32} \left ( -\frac{1}{4}\xi^{2} + 5
+ \xi +2 \sqrt{\xi^{2} +4 - 2\xi} \right )$} \, .

\centerline{$\Phi = 3\xi \sin x + b_{3}\sin 3x \quad ; \quad $
$E = \frac{\hbar \omega_{c}}{32} \left ( -\frac{1}{4}\xi^{2} + 5
- \xi +2 \sqrt{\xi^{2} +4 + 2\xi} \right )$} \, .

where 

$a_{\pm} = 2 \pm 2 \sqrt{1+\xi^2}$ \, ; \  
$ b_{0,2} = 4-\xi \pm 2\sqrt{\xi^{2}-2\xi + 4} \, ; \
b_{1,3} = 4+\xi \pm 2\sqrt{\xi^{2} + 2\xi + 4}$ \, .

A remark is in order at this stage. Since we are treating the problem 
of a point particle on circle $S^{1}$, and since the potential in the 
Schr\"odinger eq. (\ref{se}) satisfies the boundary condition
\begin{equation}
V(x+\pi) = V(x) \, ,
\end{equation}
hence physical considerations demand that the corresponding wave 
functions must also satisfy the boundary condition
\begin{equation}
\psi(x+\pi) = \psi(x) \, .
\end{equation}
In that case, the solutions obtained for $n=1,3$ are unacceptable as
they do not satisfy this boundary condition but rather they change sign
under $x \rightarrow x+\pi$. However, if we are considering it as a 
periodic problem on the full line then of course these are acceptable solutions.

We may add that the ground state energies obtained here (for $n = 0,2$) 
are useful in
another context. In particular, it has been shown using the transfer 
integral method \cite{ssf} that the classical free energy of the soliton 
bearing field theories at low temperatures is given by the ground state
energy of the corresponding Schr\"odinger like equation. Thus the ground
state energies obtained here for $n=0,2$ are of direct relevance in the
context of the classical free energy of the corresponding 
double sine-Gordon field theory.

\subsection{ Euclidean time solutions--Bounces and Instantons}
Since the effective potential also has a false vacuum, we expect that 
`bounce' solutions exist to the classical
equations of motion in Euclidean time. The classical equation
of motion in Euclidean time which we solve is given by 
\begin{equation}
{\theta '}^{2} = \omega^{2} \left [ (1-\sin \theta) (1+\sin\theta - a)
\right ] \, ,
\end{equation}
where we have added a constant to the original effective potential
in order to write it in the above form (note that this does not
effect the classical equations of motion which we intend to solve).

Here $a = {\omega_c \over \omega} < 2$. 
However, this choice of the effective potential forces us to
confine ourselves to $a<2$. For $a=2$ the equation is meaningless.
 
A straightforward
integration of the above equation 
yields the following solutions :
\be\label{t4}
\theta_{1} (\tau) = \frac{\pi}{2} + 2\tan^{-1}\left [ \frac{1}{\alpha}  
sech \sqrt{\omega(\omega-\frac{\omega_{c}}{2})} \tau \right ] \, .
\ee
\be\label{t5}
\theta_{2} (\tau) = \frac{5\pi}{2} - 2 \tan^{-1}\left [ \frac{1}{\alpha}
sech \sqrt{\omega (\omega-\frac{\omega_{c}}{2})} \tau \right ] \, .
\ee
The first of these is the bounce across the barrier at $\sin^{-1}\frac{\omega
_{c}}{2\omega}$ while the second one is across the barrier at 
$\pi - \sin^{-1} \frac{\omega_{c}}{2\omega}$. Note that the
Euclidean actions for both these solutions are the
same--this is because the barrier heights which separate the
false and the true vacua are the same. The expression for the
Euclidean action turns out to be 
\begin{equation}
S_{E} = 2mr^{2}\omega_{c}\sqrt{1-\frac{a}{2}} \left [ 
\frac{2+a}{a} - \sqrt{\frac{2-a}{2}} \sinh^{-1}\sqrt{\frac{2-a}{2}} \right ]  
\end{equation}
where $\alpha ^{2}
 = \frac{a}{2-a}$.
Using the Euclidean action one can now evaluate the decay rate of the
false vacuum by the formula $\Gamma \sim \exp \left 
(-\frac{S_{E}}{\not h}\right )$.

For a periodic potential generalization of the magnetic case problem 
one has degenerate absolute minima at ${(2n+1)\pi \over 2}$ and  
one can obtain an `instanton' which starts out at, say, $\theta
=-\frac{\pi}{2}$ (as $\tau \rightarrow -\infty$) and crosses the local minimum
at $\theta = \frac{\pi}{2}$ (as $\tau \rightarrow 0$) 
 to end up at $\theta=\frac
{3\pi}{2}$ (as $\tau \rightarrow \infty$).
To that end, let us first note that by adding a suitable constant 
one can also write the classical equation of motion in Euclidean time 
in an alternative form as
\begin{equation}
{\theta '}^{2} = \omega^{2} \left [ (1+\sin \theta) (1-\sin\theta + a)
\right ] \, .
\end{equation}
It is now easily shown that 
irrespective of whether $\omega >$ or $<\omega_c/2$, the 
instanton solution is given by 
\be\label{t6}
\theta_3 = \frac{\pi}{2} + 2\tan^{-1} \left [ \frac{1}{\sqrt{1+\frac{2\omega}{
\omega_{c}}}}\sinh \left ( \sqrt{(\omega_{c}+2\omega)2\omega} \frac{\tau}{2}
\right ) \right ] \, ,   
\ee
with the corresponding action being
\begin{equation}
S_E = -{1\over 2a^2} + {1\over 4a^3} {(4a^2-1)\over \sqrt{a^2 -1}} 
\ln \bigg [ {a+\sqrt{a^2 -1} \over a-\sqrt{a^2-1}} \bigg ] \, . 
\end{equation}
where $a = \sqrt{1+{2\omega \over \omega_c}}$.

\section{Small oscillations about instantons and bounces--
qualitative analysis}

In this section we present a qualitative analysis of the problem
of small oscillations about an instanton/bounce solution. We do {\em not}
explicitly solve the corresponding Schr\"odinger-like 
equations but discuss qualitatively 
the nature of the effective potentials, the possible existence of
negative eigenvalues and bound states. It may however be added that in 
all the cases we have been able to reduce the stability equation 
to Heun's equation. Additionally, the analysis for the small oscillations
about the DSG kink has been discussed in \cite{osc:ref}. 

The equation governing small oscillations is given by:
\begin{equation}
\left (-\partial_{\tau}^{2} + V''(\theta (\tau))\right ) \chi_{n}=
\lambda_{n} \chi_{n} \, .
\end{equation}
This is a `time--independent' Schr\"odinger equation with the potential
$V''(\theta (\tau))$. We therefore need to solve the corresponding
eigenvalue problem and look for the existence of a negative eigenvalue
which will indicate an instability. Below we write down the `potential'
in the two cases of the gravitational and magnetic fields.  

{\bf 1. : System in a gravitational field}

Since for $\omega>\omega_{0}$ we have two types of instantons we
need to evaluate $V ''(\theta)$ for each of these cases separately.

(a) Using eq. (ref{t1}), $V''(\theta)$ 
for the instanton across the $2\pi$ barrier turns out to be
\begin{equation}
V''(\theta_1 (\tau)) = -mr^{2}{\omega}^{2}(1-a^2)\frac{\left [(1-a) + (1+a)
\tanh^{4}{b\omega\tau}-6\tanh^{2}b\omega\tau \right ]}{\left [
(1-a)+(1+a)\tanh^{2}b\omega\tau \right ]^{2}} \, .
\end{equation}
This is shown in Fig 5.

(b) Using eq. (\ref{t2}), $V''(\theta (\tau ))$ for the instanton 
across the $\pi$ barrier turns out to be
\begin{equation}
V''(\theta_2 (\tau)) = -mr^{2}{\omega}^{2}(1-a^2)\frac{\left [(1-a) + (1+a)
\coth^{4}{b\omega\tau}-6\coth^{2}b\omega\tau \right ]}{\left [
(1-a)+(1+a)\coth^{2}b\omega\tau \right ]^{2}} \, .
\end{equation}
This is plotted in Fig 6.

A condition for the stability of a given instanton is the existence of
a node-less zero--mode solution (solution with a zero eigenvalue)
 to the Schr\"odinger equation governing
small oscillations. The zero mode solution can be written down very
easily-- $\chi_{0}$ is just equal to $\frac{d\theta(\tau)}{d\tau}$
(details are there is [5], [6] and [7]).

For the instanton across $\theta = 2\pi$, using eq. (\ref{t1}), one finds that 
\begin{equation}
\chi_0 = \frac{d\theta_{1}(\tau)}{d\tau} = 
(1-a^{2}) \omega \frac{1}
{[2\cosh^{2}\beta \tau -(1+a)`]}
\end{equation}
 where $\beta = \sqrt{\frac{1-a^{2}}{2}} \omega$.
Note that this solution is node-less which implies that the
corresponding instanton is stable. 

On the other hand, it can be readily shown that the instanton across
the $\theta=\pi$ barrier is also stable. 
In particular, using eq. (\ref{t2}), the zero--mode solution is given by
\begin{equation}
\chi_{0} = \frac{d\theta_{2}(\tau)}{d\tau} =  
-(1-a^{2}) \omega \frac{1}{[(1+a)
+2\sinh^{2}\beta \tau]} \, .
\end{equation}
which is clearly nodeless as well.

In the case of $\omega <\omega_0$ there is only one instanton
and using eq. (\ref{t3}), $V''(\theta)$ turns out to be
\begin{equation}
V''(\theta_3 (\tau)) = mr^2 \omega^2 \frac{\bigg [(a-1)p^4 \sinh^4 y 
+6p^2 \sinh^2 y - (a+1) \bigg ]}{(1+p^2 \sinh^2y)} \, ,
\end{equation}
and the corresponding zero-mode solution is given by
\begin{equation}
\chi_0 = \frac{d\theta_{3}(\tau)}{d\tau} = \frac{2\omega_0 \cosh y}
{1+p^2 \sinh^2 y} \, , 
\end{equation}
which is clearly node-less. Here $y = {\omega_0 \tau \over p}$ with 
$a= {\omega_0^2 \over \omega^2}$ and $p = \sqrt{{a \over a-1}}$.

{\bf Case 2: System in a magnetic field}

For the bounce solution in the PORC in a magnetic field, 
using eq. (\ref{t4}) (or (\ref{t5})), 
the potential $V''(\theta)$ in the Schr\"odinger-like 
equation turns out to be
\begin{equation}
V''(\theta_{1,2}(\tau)) =
\left (\frac{mr^2{\omega}^2}{2-a} \right ) 
\frac{\left [-a^2\cosh^{4}\beta \tau +
8a\cosh^{2}\beta \tau - (4-a^2) \right ]} 
{(1+ \alpha^2 \cosh^{2}\beta \tau)^2} \, .
\end{equation}
This is plotted in Fig.7.  
The corresponding zero-mode solution is given by
\begin{equation}
\chi_0 = \frac{d\theta_{1,2}(\tau)}{d\tau} = \alpha \omega \sqrt{2(2-a)}
\left (\frac{\sinh{y}}{1+ \alpha^2 \cosh^2{y} } \right ) \, ,  
\end{equation}
which clearly has a node at $\tau=0$. Here $y = {(2-a)\omega \tau \over 2}$.

On the other hand, corresponding to the instanton solution in a magnetic 
field, using eq. (\ref{t6}), $V''(\theta)$ 
turns out to be
\begin{equation}
V''(\theta_{3}(\tau) = {m r^2 \omega^2 (a+1) \over a} 
{\bigg [a(6+a) -2(1+3a) \cosh^2{y} + \cosh^4{y} \bigg ] 
\over (a + \cosh^2{y})^2} \, .
\end{equation}
The corresponding zero-mode solution is given by 
\begin{equation}
\chi_0 = \frac{d\theta_{3}(\tau)}{d\tau} \propto 
\left ({\cosh{y} \over a + \cosh^2{y}} \right ) \, ,  
\end{equation}
which clearly is node-less. Here $a = {2\omega \over \omega_c}, \ y = 
\sqrt{a(a+1)} {\omega_c \tau \over 2}$.

\section{Summary and outlook}

We have discussed a variety of aspects of the toy model of a particle
on a rotating loop. After defining a generic class of models (via a
choice of a class of trigonometric potentials )
 we identified
two specific cases. These included the effects of a (i) uniform gravitational
field and (ii) a uniform magnetic field. The classical mechanics for both
these cases was worked out. Subsequently we dealt with the quantum 
mechanical problem in either case from the 
(i) exact as well as the (ii) semi-classical
standpoints. For the gravitational case we found exact solutions by
identifying the time--independent Schr\"odinger equation with the
Whittaker--Hill equation. These solutions were {\em formal} in the sense
that the energy eigenvalues remain hidden in the continued fractions.
We have not attempted to arrive at the expressions for the energy
eigenvalues by solving the continued fractions. Hence this discussion
appears in the Appendix to the paper. Apart from the 
exact solutions we have also looked at the instantons in this
model. We constructed the instantons and the corresponding Euclidean
actions which gave us a feeling for the quantum tunneling phenomena
in this example. In the magnetic case the potential
also has a false vacuum. Further, for special values of the parameters
we were able to obtain some of the energy eigenvalues and eigenfunctions. 
Thereafter,
we constructed the corresponding bounce (instanton) 
solutions and their Euclidean
actions to arrive at the decay rate of the false vacuum (feel 
for quantum tunneling phenomena). We then 
looked into the problem of small oscillations about the instantons and bounces
by investigating the nature of the potential in the Schr\"odinger--type
equation which govern these perturbations. This section is largely 
qualitative. Even though we did not solve the relevant Schr\"odinger-like 
equation, we have shown that in all the cases the Schr\"odinger equation 
reduces to Heuns equation which is known to have four regular singular points.
Further, as expected we have seen that the zero mode is stable (unstable) 
depending on whether one is considering an instanton (bounce) solution.
Finally, the connection of these toy examples
 with several models in field
theory and statistical mechanics was pointed out. 

The aim of the paper has been to point out the diverse aspects of the 
problem of a particle on a rotating loop. It is quite illuminating
that a lot of important concepts in field theory  can be illustrated
through the study of this simplistic example. We conclude by pointing
out some other aspects which may be dealt with in future.

\begin{itemize}
\item{ Aspects of chaos in PORC with a $\delta$ function kick--the 
generalization of the kicked pendulum/kicked rotor}
\item{Investigating what happens at finite temperature by utilizing techniques
of finite temperature quantum mechanics/quantum field theory}
\item{Exact classical and quantum solutions for the model with a gravitational,
an electric and a magnetic field all put in together}  
\end{itemize}

\section*{Acknowledgements}
SK thanks IUCAA, Pune for financial support during the period when this
work began. 

\newpage

\centerline{{\bf APPENDIX : Exact solutions for PORC in a gravitational field}}
\vspace{.2in}

The time-independent Schr\"{o}dinger
equation for this system is given by
\begin{equation}
\frac{d^{2}\psi}{d\theta^{2}}+\frac{2m}{\hbar^{2}}\left[
E+\frac{1}{4}mr^{2}\omega^{2}-mgr\cos\theta-\frac{1}{4}mr^{2}\omega^{2}
\cos2\theta\right]\psi=0 \, .
\end{equation}
Introducing a new variable $\alpha=\theta/2$ we arrive at the
following equation
\be\label{c3}
\frac{d^{2}\psi}{d\alpha^{2}}+\left[\left(\frac{8mE}{\hbar^{2}}+
\frac{2m^{2}r^{2}\omega^{2}}{\hbar^{2}}\right)-
\frac{8m^{2}gr}{\hbar^{2}}\cos2\alpha-\frac{2m^{2}r^{2}\omega^{2}}
{\hbar^{2}}\cos4\alpha\right]\psi=0 \, .
\ee
This equation is the Whittaker-Hill equation.We
now digress briefly to a discussion of the solutions of this
equation.

In the theory of periodic differential equations, the
Whittaker-Hill equation is a special case of the general Hill
equation given by 
\begin{equation}
\frac{d^{2}\psi}{d\theta^{2}}+\left(\sum_{n=0}^{\infty}A_{n}\cos
2n\theta\right)\psi=0 \, .
\end{equation}
Whittaker-Hill equation is a special case of this general Hill equation
and is also known as the three term Hill equation, i.e. we have
only the $n=0,1,2$ terms remaining in the infinite series given above.
Thus we have
\begin{equation}\label{8.1}
\frac{d^{2}\psi}{d\theta^{2}}+\left(A_{0}+A_{1}\cos2\theta+
A_{2}\cos4\theta\right)\psi=0 \, .
\end{equation}
Such equations arise when the Helmholtz equation
$\left(\nabla^{2}+k^{2}\right)
\psi=0$ is separated in a general paraboloidal coordinate system.
We shall be concerned here with the form of equation for which 
$A_{1}$ and $A_{2}$ are both negative.Following \cite{whe:ref} we define 
\begin{equation}
|A_{2}|=\frac{1}{8}\omega^{2},A_{0}=\eta+\frac{1}{8}\beta^{2},
A_{1}=-\rho\beta \, .
\end{equation}
Hence we have
\be\label{c4}
\frac{d^{2}\psi}{d\theta^{2}}+\left(\eta+\frac{1}{8}\beta^{2}-
\rho\beta\cos2\theta-\frac{1}{8}\beta^{2}\cos4\theta\right)\psi=0 \, .
\ee
The transformation
$\psi=\phi\exp(-\frac{1}{4}i\beta\cos2\theta)$ reduces (\ref{8.1}) to
Ince's equation
\begin{equation}
\frac{d^{2}\phi}{d\theta^{2}}+i\beta\sin2\theta\frac{d\phi}{d\theta}
+\left[\eta+\beta(\rho+i)\cos2\theta\right]\phi=0 \, .
\end{equation}
Solutions to the above equation falls into four classes depending on whether
the function $\phi$ is even or odd and has period $\pi$ or
$2\pi$.  These solutions are in terms of infinite Fourier series
with the relation between the coefficients of the series given
in terms of three term recurrence relations. The recurrence
relations can be reduced to an infinite continued fraction which
is known as the corresponding characteristic equation for the
relevant function. We shall primarily be concerned with
solutions of period $\pi$. This is because 
the equation governing our physical problem is written in terms
of a new variable $\alpha=
\theta/2$. Thus $\pi$ periodic solution in $\alpha$ will turn out to be
 $2\pi$ periodic solutions in $\theta$. We now list the even and odd $\pi$ 
 periodic solutions of the Whittaker-Hill equation and the relevant 
 continued fractions. Details about how these are obtained and a
  more thorough treatment of various issues related to periodic
   differential equations can be found in 
\cite{whe:ref}.
\subsection{Even Solutions of Period $\pi$}
\begin{equation}
\phi(\theta)=\sum_{r=0}^{\infty}A_{2r}\cos2r\theta\equiv gc_{2n}(\theta,
\omega,\rho) \, ; \ A_{2r} = X_{2r}B_{2r} \, ,
\end{equation}
where
\begin{equation}
X_{2r}=(-\rho+i)(-\rho+3i)\cdots\left[-\rho+(2r-1)i\right] \, ; 
r\geq1 \, , \ X_{0}=1 \, , 
\end{equation}
and
\begin{equation}
\frac{B_{2r}}{B_{2r+2}}=\frac{\frac{1}{2}\beta}{4r^{2}-\eta-}
\frac{\frac{1}{4}\beta^{2}\{\rho^{2}+(2r+1)^{2}\}}{4(r+1)^{2}-\eta-}
\frac{\frac{1}{4}\beta^{2}\{\rho^{2}+(2r+3)^{2}\}}{4(r+2)^{2}-\eta-}
\cdots \, .
\end{equation}
The Continued fraction for characteristic values is given by
\be\label{c1}
-\frac{2\eta}{\beta(\rho^{2}+1)}=\frac{\beta}{4-\eta-}
\frac{\frac{1}{4}\beta^{2}(\rho^{2}+9)}{16-\eta-}
\frac{\frac{1}{4}\beta^{2}(\rho^{2}+25)}{36-\eta-}\cdots \, .
\ee
\subsection{Odd Solutions of Period $\pi$}
\begin{equation}
\phi(\theta)=\sum_{r=1}^{\infty}C_{2r}\sin2r\theta\equiv gs_{2n+2}(\theta,
\omega,\rho) \, ; \ C_{2r} = Y_{2r} D_{2r} \, ,
\end{equation}
where
\begin{equation}
Y_{2r}=(-\rho+3i)(-\rho+5i)\cdots\left[-\rho+(2r-1)i\right] \, , \ 
r\geq2 \, , \ Y_{2}=1 \, , 
\end{equation}
and
\begin{equation}
\frac{D_{2r}}{D_{2r-2}}=\frac{\frac{1}{2}\beta}{4r^{2}-\eta-}
\frac{\frac{1}{2}\beta^{2}\{\rho^{2}+(2r+1)^{2}\}}{4(r+1)^{2}-\eta-}
\frac{\frac{1}{2}\beta^{2}\{\rho^{2}+(2r+3)^{2}\}}{4(r+2)^{2}-\eta-}
\cdots \, .
\end{equation}
The Continued fraction for characteristic values is given by
\be\label{c2}
\frac{4-\eta}{\frac{1}{2}\beta(\rho^{2}+9)}=\frac{\frac{1}{2}\beta}{16-\eta-}
\frac{\frac{1}{4}\beta^{2}(\rho^{2}+25)}{36-\eta-}
\frac{\frac{1}{4}\beta^{2}(\rho^{2}+49)}{64-\eta-}\cdots \, .
\ee
Equations (\ref{c1}) and (\ref{c2}) are the equations which will give us the
energy levels. Comparing eqs. (\ref{c3}) and (\ref{c4}) we find that
\begin{equation}
\eta=\frac{8mEr^{2}}{\hbar^{2}} \, , \ \frac{1}{8}\beta^{2}
=\frac{2m^{2}r^{4}\omega
^{2}}{\hbar^{2}} \, , \ 
\rho=\frac{2mgr}{\hbar\omega} \, .
\end{equation}
Fixing $m,r$ and $\omega$ we can evaluate the energy levels of
the system by solving the respective characteristic equations
(for even and odd functions) iteratively.


\begin{references}
\bibitem{gold:cm} H. Goldstein, {\em Classical Mechanics} (Addison Wesley,
Reading, 1985) p. 68 ; V. I. Arnold, {\em Mathematical methods of
classical mechanics} (Springer, Berlin, 1978) p.87.
\bibitem{porc:ajp} T. Bernstein, A Mechanical Model of the Spin-Flop 
Transition in Antiferromagnets, Am. Jr. Phys. {\bf 39}, 832-834 (1971);
R. Alben, An Exactly Solvable Model Exhibiting A Landau Phase Transition, 
Am. Jr. Phys. {\bf 40}, 3-8 (1972); E. Guyon, Second-Order Phase 
Transitions: Models and Analogies, Ame. Jr. Phys.
{\bf 43}, 877-881 (1975); J. Sivardiere, A Simple Mechanical Model 
Exhibiting a Spontaneous Symmetry Breaking, Am. Jr. Phys. {\bf 51}, 
1016-1020 
(1983); J. R. D. de Felicio and O. Hipolito, Spontaneous Symmetry 
Breaking in a Simple Mechanical Model, Am. Jr. Phys.
{\bf 53}, 690-693 (1985); J. E. Drumheller, D. Raffaelle and M. Baldwin
, An Improved Mechanical Model to Demonstrate the First- and Second-Order
Phase Transitions of the Easy-Axis Heisenberg Anti-ferromagnet, 
Am. Jr. Phys. {\bf 54}, 1130-1133 (1986); E. Marega Jr., S. C. Zillo
and L. Ioriatti, Electromechanical Analog for Landau's Theory of 
Second-Order Symmetry Breaking Transitions, 
Am. Jr. Phys. {\bf 58}, 655-659 (1990).
\bibitem{flet:ajp97} G. Fletcher, A Mechanical Analog of First- and 
Second-Order Phase Transitions, Am. Jr. Phys. {\bf 65}, 74-80 (1997).  
\bibitem{sk:pla92} S. Kar, An Instanton Approach to Quantum Tunneling 
For a Particle on a Rotating Circle, Phys. Letts. {\bf A 168}, 179-186 (1992).
\bibitem{gen:inst} S. Coleman, {\it Aspects of Symmetry} (Cambridge
University Press, Cambridge, 1985) Ch. 7.
\bibitem{gen1:inst} R. Rajaraman, {\em Solitons and Instantons} (North-
Holland, Amsterdam, 1982) Ch. 10.
\bibitem{kleinert:pi} H. Kleinert, {\it Path Integrals in Quantum
Mechanics, Statistics and Polymer Physics} (World Scientific, Singapore,
1990).
\bibitem{hks} A. Khare, S. Habib and A. Saxena, Exact Thermodynamics of the
Double Sinh-Gordon Theory in 1+1 Dimensions, Phys. Rev. Lett. {\bf 79}
,3797 (1997); S. Habib, A. Khare and A. Saxena, Statistical Mechanics of 
Double Sinh-Gordon Kinks, Physica {\bf D123} 341-356 (1998).
\bibitem{km} A. Khare and B.P. Mondal, Anti-Isospectral Transformations, 
Orthogonal Polynomials and Quasi-Exactly Solvable Problems, 
J. Math. Phys. {\bf 39}, 3476-3486 (1998). 
\bibitem{raz:ajp} M. Razavy, An Exactly Soluble Schr\"odinger Equation With 
a Bistable Potential, Ame. Jr. Phys. {\bf 48}, 285-288 (1980);
 A. Ushveridze, {\it Quasi-Exactly Solvable Models in Quantum Mechanics} 
(Inst. of Phys. Publishing, Bristol, 1994).
\bibitem{kp:ajp} T. Pradhan and A. Khare, Plane Pendulum in Quantum 
Mechanics, Am. Jr. Phys. {\bf 41}, 59-66 (1973).
\bibitem{leu} K.M. Leung, Mechanical Properties of Double-Sine-Gordon 
Solitons and the Application to Anisotropic Heisenberg Ferromagnetic 
Chains, Phys. Rev. {\bf B27}, 2877-2888 (1983).
\bibitem{ptk} R. Pandit, C. Tannous and J.A. Krumhansl, Statistical Mechanics 
of a Classical One-Dimensional Canted Anti-ferromagnet II. Solitons, 
Phys. Rev. {\bf B28},
289-299 (1983).
\bibitem{boc} R.K. Bullough and P.J. Caudrey, in {\it Nonlinear Evolution 
Equations Solvable by the Inverse Spectral Transform}, edited by F. Calogero
(Pitman, London, 1978) pp. 180-224.
\bibitem{cgm} C.A. Condat, R.A. Guyer and M.D. Miller, Double Sine-Gordon 
Chains, Phys. Rev. {\bf B27},
474-494 (1983).
\bibitem{det} R.M. DeLeonardis and S.E. Trullinger, Classical Statistical
Mechanics of One Dimensional Polykink Systems, Phys. Rev. {\bf B27}, 
1867-1886 (1983).
\bibitem{mat} K. Maki and T. Tsuneto, Magnetic Resonance and Spin Waves in the 
A-Phase of Superfluid  $^3 He$, Phys. Rev. {\bf B11}, 2539-2543 (1975); 
K.Maki and P. Kumar, Creation of Magnetic Solitons in Superfluid $^3 He$, 
Phys. Rev. {\bf B14}, 3920-3928 (1976) ; Magnetic Solitons in 
Superfluid $He^3$,
Phys. Rev. {\bf B14}, 118-127 (1976).
\bibitem{buc} R.K. Bullough and P.J. Caudrey, in {\it Proceedings of the 
Fourth Rochester Conference on Coherence and Quantum Optics}, edited by 
L. Mandel and E. Wolf (Plenum, New York, 1978).
\bibitem{dbc} S. Duckworth, R.K. Bullough, P.J. Caudrey and J.D. Gibbon, 
Unusual Soliton Behaviour in the Self-Induced Transparency of Q(2) 
Vibration-Rotation Transitions, Phys. Lett. {\bf 57A}, 19-22 (1976).
\bibitem{rem} M. Remoissenet, Nonlinear Excitations in a Compressible Chain 
of Dipoles, J. Phys. {\bf C14}, L335-L338 (1981).
\bibitem{hls} A.J. Hopfinger, A.J. Lewanski, T.J. Sluckin and P.L. Taylor, 
in {\it Soliton and Condensed Matter Physics}, edited by A.R. Bishop and 
T. Schneider (Springer, Berlin, 1978) pp. 330.
\bibitem{col} S. Coleman, Quantum Sine-Gordon Equation as the Massive 
Thirring Model, Phys. Rev. {\bf D11} 2088-2097 (1976).
\bibitem{bha} S. Li, D.B. Mitchell and R.K. Bhaduri, A Modified Massive 
Thirring Model and New Solitonic Coats on the (1+1)-Dimensional Chiral Bag,
Ann. of Phys. {\bf 193} 93-101 (1989).
\bibitem{gr} I.S. Gradshteyn and I.M. Ryzhik, {\it Table of Integrals, 
Series and 
Products} (Academic Press, New York, 1980) pp. 245.
\bibitem{ssf} D.J. Scalapino, M. Sears and R.A. Ferrell, Statistical Mechanics 
of One-Dimensional Ginzburg-Landau Fields, Phys. Rev. {\bf B6},
3409-3416 (1972).
\bibitem{osc:ref} P. Sodano, M. El--Batanouny, C. R. Willis, 
Eigenfunctions of the Small Oscillations about the Double-Sine-Gordon Kink, 
Phys. Rev.
{\bf B 34}, 4936-4939 (1986). 
\bibitem{whe:ref} K. Urwin and F. M. Arscott, Theory of the
Whiitaker--Hill Equation, Proc. Roy. Soc. (Edin)
, 28 (1970); K. Urwin, Perturbation solutions of the Whittaker Hill
Equation, Jr. Inst. Math. Appl. {\bf 3}, 169 (1967); F. M.
Arscott, Periodic Differential Equations (Oxford, Pergamon Press, 1964) 
\end{references}
\end{document}